# Prediction of heart rate response to conclusion of spontaneous breathing trial by fluctuation dissipation theory

[1]Man Chen, [1]Liang Ren Niestemski, [2]Robert Prevost, [2]Michael McRae, [3]Sharath Cholleti, [3]Gabriel Najarro, [3]Timothy G. Buchman, and [1,2]Michael W. Deem

[1] Physics & Astronomy, Rice University, Houston, TX

[2] Bioengineering, Rice University, Houston, TX

[3] Emory University, Atlanta, GA

**Abstract**

The non-equilibrium fluctuation dissipation theorem is applied to predict how critically ill patients respond to treatment, based upon data currently collected by standard hospital monitoring devices. This framework is demonstrated on a common procedure in critical care: the spontaneous breathing trial. It is shown that the responses of groups of similar patients to the spontaneous breathing trial can be predicted by the non-equilibrium fluctuation dissipation approach. This mathematical framework, when fully formed and applied to other clinical interventions, may serve as part of the basis for personalized critical care.

**Introduction**

Biological health is not a single, stable, fixed point. Rather, health reflects a rich interplay of complex dynamics. Recent observations suggest that erosion of the mechanisms that underlie natural physiological complexity may be one of the most significant damaging effects of trauma or illness [1]. For example, loss of heart rate variability is associated with deterioration of health, and loss of natural correlations in gait is associated with postural imbalance. To preempt that erosion, there is an unmet need for predictive physiology [2]. Currently, clinical predictions are based on pattern classification. By echoing predictive meteorology–that is, the use of dense data, high-speed computing, and repeated application of simple physical laws–we will make an important step in developing a framework for predictive physiology.

In this paper, we apply the conceptual framework of fluctuation-dissipation theory to predict the response to a common clinical intervention from historical fluctuations, data that are characteristic of all physiologic time series data. With this theory, significantly less data are required to make predictions compared to a general data fitting approach. The non-equilibrium fluctuation dissipation theorem (FDT) relates the response of a system to a perturbation to the fluctuations in the stationary state of the system. We seek to apply the FDT to human physiology. Our model intervention is the spontaneous breathing trial (SBT), a frequent procedure in critical care during which mechanical ventilation is briefly suspended while the patient breathes for a time without that support. Assisted breathing is a common intervention: one day point-prevalence study showed that 40% (20,60) (25th/75th percentile) of ICU patients in in North America receive mechanical ventilation on a typical day [3]. The SBT is used to assess the readiness of patients for removal from assisted breathing. The SBT is also a stress upon the other organs of the patient, in particular upon the heart. Thus, we characterize the clinical stress, the SBT, as a perturbation of heart rate dynamics. The FDT allows us to predict the heart rate recovery after the SBT stress. Since the rate of recovery from SBT stress is believed to be an indicator of the capacity for liberation from mechanical physiological supports, such a prediction is clinically significant.

We here use clinical data from Emory University Hospital to test the far-from-equilibrium fluctuation-dissipation theory (FDT) predictions for the response of the heart rate to the SBT.

Initiation of the SBT is the start of stress upon the heart of the patient.  Similarly, termination of the SBT and continuation of the assisted breathing is the end of the stress upon the heart.  We seek to predict how the heart rate responds to the stress ending at the termination of SBT.

An extension of the usual fluctuation-dissipation theorem from equilibrium to non-equilibrium but steady-state systems is necessary to study physiology. Because various physiological parameters are measured over time for patients in intensive care units, a fully formed theory could help form the foundation for more individualized critical care, and, in turn, for better patient outcomes.  In this project, we propose to develop a powerful tool in the mathematics of fluctuations for predicting the patient response to interventions to the non-equilibrium states so often seen in the clinical physiology of trauma and illness.  Introduction of the non-equilibrium fluctuation-dissipation theorem to physiologic systems is a creative idea for analyzing and predicting the dynamics of patient response. It is a fresh approach to individualized medicine. By developing this generalized theorem, we do away with the assumption made in our prior work that the data are Gaussian, and instead compute the probabilities and correlations directly from the data sets, which we fully expect will improve the accuracy of the prediction.  Moreover, by applying our extended theorem, we will be able to make predictions not only about heart rate and blood pressure, but ultimately about multiple physiological variables.  Especially exciting to us is the expectation that application of our extended theorem may make it possible to detect the onset of transitions from the disease or ill state back to health.

The mathematical novelty lies in two domains:  1) It connects a mathematics concept (FDT) with a clinical event (SBT) and thereby elucidates both.  2) It moves from equilibrium assumptions to a more robust non-equilibrium FDT. Both novel aspects are broadly applicable to clinical medicine and especially to the traumatic injuries and critical illnesses that are decidedly far from equilibrium.

The equilibrium fluctuation dissipation theorem was first proved by Callen and Welton in 1951 [4]. The theorem is a fundamental result for systems near thermodynamic equilibrium and provides a general relation between microscopic fluctuations and macroscopic responses. Applications to calculation of transport coefficients were realized early on [5].  Modern applications include predictions of responses for networked systems [6].   The intuition is that responses to perturbations are predictable from equilibrium fluctuations, because fluctuations and responses occur at equilibrium. The generalization of this theory to non-equilibrium systems, such as physiology, is non-trivial [7]. From this generalized theorem, we can obtain the relationship between the event-to-event microscopic fluctuations (HRV) of physiology during a stress and the macroscopic response (HRR) after the stress.

The FDT has previously been applied to the SBT.  In that previous approach, we assumed the dynamics of the heart rate follows a discrete-time Markov process with no memory.  We also assumed the data were distributed according to a Gaussian process.  These two assumptions meant that the predicted response function was an exponentially decaying function of time.  Many data sets could be analyzed with these assumptions.  Some data sets, however, showed non-exponential decay.  We here apply the generalized FDT [7], which does not required these assumptions.

In the present work, we classify the heart rate dynamics of individuals into a limited number of categories.  We predict the cardiac response to the SBT of each group of the categories using data from all members of each group.  That is, we classify the patients, and we predict the average response of each class to the SBT. We describe details of our prediction in the Methods section.  We review the FDT and how it is applied to predict the heart rate response to SBT.  We describe the classification of the patients.  We present the results of the calculations in the Results section. We describe how the data were collected and the groupings of the patients.  We discuss the ability of the FDT to predict the heart rate response in the Discussion section.  We discuss some of the practical data collection issues and their impact upon our ability to make predictions.  We end with a Conclusion section.

**Methods**

The generalization of the FDT builds upon the Jarzynski equation [8]. Jarzynski gives the equilibrium free energy difference, ΔF, between two configurations A and B in terms of an ensemble average of finite-time measurements of the work, W, performed on the systems as it is switched from A to B:

$$\overline{\exp(-\beta W)} = \exp(-\beta \Delta F) \qquad (1)$$

The equation is remarkable, as it had previously been assumed that free energies are computable only from reversible work measurements. Based on Jarzynski's derivation, Hatano and Sasa [9] studied Langevin dynamics for a system with a set of time-dependent parameters, $\lambda_\alpha(t)$, describing non-equilibrium steady state and derived the following result:

$$\left\langle \exp\left\{ -\int_{t_i}^{t_f} dt\, \dot{\lambda}_\alpha(t) \frac{\partial \phi(c(t); \lambda(t))}{\partial \lambda_\alpha} \right\} \right\rangle = 1 \qquad (2)$$

Here the potential $\phi(x;\alpha) = -\log \rho_{ss}(x;\alpha)$, $\rho_{ss}(x;\alpha)$ is the probability distribution function at steady state, and $\lambda_\alpha$ is the set of control parameters for the system. This non-equilibrium generalization allowed Prost et al. in 2009 [7] to generalize the FDT to non-equilibrium systems evolving between two steady states and following Markovian dynamics:

$$< \frac{\partial \phi(c(t); \lambda^{ss})}{\partial \lambda_\alpha} > = \int_{ti}^{t} \chi_{\alpha\gamma}(t-t')\delta\lambda_\gamma(t')dt'$$

$$x_{\alpha\gamma}(t-t') = \frac{d}{dt}C_{\alpha\gamma}(t-t') = \frac{d}{dt}\left\langle \frac{\partial \phi(c(t); \lambda^{ss})}{\partial \lambda_\alpha} \frac{\partial \phi(c(t'); \lambda^{ss})}{\partial \lambda_\gamma} \right\rangle_{ss} \qquad (3)$$

For our application, the control parameter will be initiation and termination of the SBT. This remarkable theorem is also obtained under more general conditions with steady state variables and is valid for non energy conserving dynamics. Theorem (3) holds irrespective of the existence of nonlinearities and of the spatial dimension. This theorem is the basis for our prediction of patient heart rate responses to the SBT. Further generalizations to non-stationary systems are possible [10, 11].

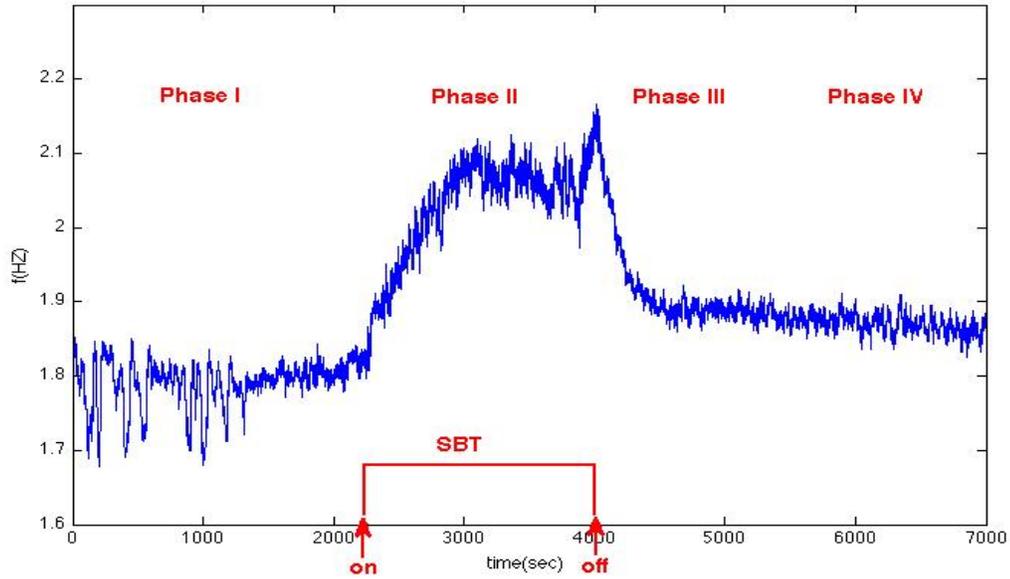

**Figure 1.** Heart rate of one patient before, during, and after the SBT. We divide the whole duration of the data set into four time intervals, or "phases": phase I, from the beginning of the data set to the beginning of SBT; phase II, SBT; phase III, relaxation of the heart rate to a new steady state; and phase IV, from the moment the heart rate becomes stationary again to the end of the data set.

We split the data into four intervals. Phase I is before the SBT. Phase II is the SBT. Phase III is the response to the SBT. Phase IV is after the patient has recovered from the SBT. These phases are shown in Figure 1. We seek to predict the response to SBT, phase III. We make this prediction for a group of patients.

We use Eq. (3) by assuming that the termination of the SBT is equivalent to a step function perturbation. We further assume that the observable is the heart rate, $\frac{d\phi}{d\lambda} = f$, where $f$ is the heart rate. Eq. 3 then states

$$<f>^{IIIp}(t) = f(t_0) + (f(t_0) - <f>^{IV}) \frac{<c(0)-c(t-t_0)>^{II}}{c(0)^{II}} \qquad (4)$$

Here $c(t-t_0) = <f(t)-<f>^{II}><f(t_0)-<f>^{II}>$ is the autocorrelation function of the steady state. For an individual patient, i.e. a patient with index $i$, we can predict the heart rate of Phase III using the fluctuations in Phase II, simply following

$$<f>_i^{IIIp}(t) = f_i(t_0) + (f_i(t_0) - <f>_i^{IV}) \frac{<c_i(0)-c_i(t-t_0)>^{II}}{c_i(0)^{II}}$$

We now generalize equation (4) to predict the average response of a group of patients. The correlation function will thus contain the fluctuations in the group, in addition to fluctuations of the data within one patient. There are a number of details to implement this formula for a group prediction. These details are described below in (i) Scaling and group average (ii) The full cross correlation method:

(i) **Scaling and group average**
We define the heart rate at the starting period of phase III to be $f^0$. The start times of phase II, III and IV are $t_0^{II}$, $t_0^{III}$, and $t_0^{IV}$, respectively. The initial heart rate $f^0$ is calculated as the average value of the heart rate $f$ from $t_0^{III} - 10\Delta t$ to $t_0^{III} + 10\Delta t$, where $\Delta t$ is the inter heart beat time step. The steady state value of the heart rate in phase IV, $<f>^{IV}$, is the average of the heart rate data in phase IV. For each patient I, where i is the patient index, we scale the heart rate data following Equation (5),

$$\bar{f}_i = \frac{f_i - <f_i>^{IV}}{f_i^0 - <f_i>^{IV}} \qquad (5)$$

Assuming that we have n patients in one group with index numbers i=1,2,…,n, then the normalized average heart rates among all patients in the group for phase II and III are

$$<\bar{f}_i>^{II} = \frac{1}{n}\sum_{i=1}^{n}\frac{f_i^{II} - <f_i>^{IV}}{f_i^0 - <f_i>^{IV}} \qquad (6)$$

$$<\bar{f}_i>^{III} = \frac{1}{n}\sum_{i=1}^{n}\frac{f_i^{III} - <f_i>^{IV}}{f_i^0 - <f_i>^{IV}} \qquad (7)$$

Since different patients have different lengths of phase II and phase III within a group, the phase II and III data must be truncated to the smallest data set in order to find the average heart rates among all patients in that group. We define $t_{min}^{II}$ and $t_{min}^{III}$ as the shortest time span of phase II and phase III, respectively, among all patients within the group. The normalized average heart rates $<\bar{f}_i>^{II}$ (from $t_0^{III} - t_{min}^{II}$ to $t_0^{III}$) and $<\bar{f}_i>^{III}$ (from $t_0^{III}$ to $t_0^{III} + t_{min}^{III}$) are shown in Fig 2 (blue curve)

(ii) **The full cross correlation method**
In order to make a prediction of phase III heart rate $<f>^{IIIp}$ of the group, we follow equation (4) with the same scaling method as in (5).

$$<\bar{f}>^{IIIp}(t + t_0^{III}) = \frac{<\delta\bar{f}(t')\delta\bar{f}(t'+t)>^{II}}{<\delta\bar{f}(t')^2>^{II}} \text{ where } \delta\bar{f} = \bar{f} - <\bar{f}>, t' \in (t_0^{II}, t_0^{III} - t). \qquad (8)$$

When calculating the autocorrelation of phase II data on the right hand side (RHS) of equation (8), we utilize all the patient data in the group. Specifically, there are $n^2$ pairs of patient correlation functions. To find the correlation of heart rates for patients *i* and *j*, we define $t'^{II}_{min}$ as the shortest time span of phase II of the (i,j) pair, and $t'^{III}_{min}$ as the shortest time span of phase III of the (i,j) pair. We compute the autocorrelation on the RHS of equation (8) for all the $n^2$ pairs in this way. The prediction of the scaled average for the group, $<\bar{f}>^{IIIp}(t)$, is

$$<\bar{f}>^{IIIp}(t + t_0^{III}) = \frac{1}{n^2}\sum_{i=1}^{n}\sum_{j=1}^{n}\frac{\frac{1}{t'^{II}_{min}-t+\Delta t}\sum_{t'=t_0^{II}}^{t_0^{II}+t'^{II}_{min}-t}\delta\bar{f}_i^{II}(t')\delta\bar{f}_j^{II}(t+t')}{\frac{1}{t'^{II}_{min}}\sum_{t''=t_0^{II}}^{t_0^{II}+t'^{II}_{min}}\delta\bar{f}_i^{II}(t'')\delta\bar{f}_j^{II}(t'')} \qquad (9)$$

We show the average prediction of phase III, $<\bar{f}>^{IIIp}(t)$ utilizing the full cross correlation method among all the patients in the group in Figure 2 (black curve).

**Theory for grouping patients**

In order to obtain more accurate predictions for the heart rate recovery (HRR) after the SBT physiological stress, we group patients into categories and average their heart rate profiles. Averaging signals among patients with similar characteristics not only improves signal-to-noise ratio of the heart rate signal, but also improves predictions for heart rate recovery in phase III by utilizing more data in the correlation calculation. Since cardiopulmonary fitness varies among patients, we group them according to the following SBT heart rate characteristics: mean slope of the HRR prediction, maximum heart rate achieved in phase II and phase III, and minimum heart rate achieved in phase II and phase III. The mean slope of the HRR predictions of patients is calculated by taking the average finite difference of HRR prediction, $<\bar{f}>_i^{IIIp}(t)$, divided by the finite difference of time t, *i.e.* $\Delta t$.

$$<\frac{d<\bar{f}>_i^{IIIp}(t)}{dt}> = \frac{1}{(t_0^{IV}-t_0^{III})/\Delta t}\sum_{t=t_0^{III}}^{t_0^{IV}}\frac{<\bar{f}>_i^{IIIp}(t+\Delta t)-<\bar{f}>_i^{IIIp}(t)}{\Delta t} \qquad (10)$$

The maximum and minimum phase II and III values are obtained by finding the maximal and minimal data points in those intervals. The criteria for grouping and the resulting patient groups are shown in Tables 1 and 2.

Table 1: Metrics for patient grouping sorted by mean HRR slope

| Patient | Mean HRR slope (BPM/s) | Maximum heart rate (BPM) | Minimum heart rate (BPM) |
|---|---|---|---|
| 2 | -0.0294 | 148.45 | 93.51 |
| 35 | -0.0251 | 84.46 | 59.26 |
| 4 | -0.0220 | 107.46 | 90.00 |
| 92 | -0.0215 | 84.21 | 57.42 |
| 7 | -0.0187 | 146.92 | 103.05 |
| 128 | -0.0186 | 153.18 | 80.49 |
| 29 | -0.0162 | 146.85 | 96.14 |
| 55 | -0.0128 | 107.46 | 80.04 |
| 91 | -0.0127 | 74.2268 | 49.88 |
| 96 | -0.0124 | 100.68 | 86.18 |
| 75 | -0.00996 | 109.92 | 69.07 |
| 8 | -0.00921 | 121.99 | 91.72 |
| 66 | -0.00908 | 115.18 | 87.80 |
| 68 | -0.00787 | 92.08 | 70.94 |
| 122 | -0.005.56 | 96.64 | 77.42 |
| 90 | -0.00539 | 125.22 | 100.70 |
| 1 | -0.00398 | 117.02 | 91.15 |
| 115 | -0.00253 | 130.26 | 98.63 |
| 119 | -0.00243 | 90.49 | 82.29 |
| 9 | -0.00851 | 105.02 | 92.95 |
| 74 | -0.000426 | 89.93 | 70.94 |
| 3 | 0.00 | 122.84 | 99.38 |
| 93 | 0.00 | 81.82 | 57.79 |

Table 2: Patient groups corresponding to error analyses in Figure 1 (a-h)

| Patient Group | Patient numbers | Grouping Criterion | Criterion Value |
|---|---|---|---|
| a | 2,4,7,8,29,35,55,66,68,75,91,92,96,128 | Mean HRR slope | < -0.005 BPM/s |
| b | 1,90,115,119,122 | Mean HRR slope | > -0.005 BPM/s |
| c | 35,92 | Mean HRR slope, maximum heart rate, and minimum heart rate | - |
| d | 2,35,92,128 | Mean HRR slope | < -0.019 BPM/s |
| e | 1,4,29,75,92,96,115,122 | Random assignments | - |
| f | 2,7,55,119 | Random assignments | - |
| g | 4,35,92 | Group c plus one additional patient, whose mean HRR slope falls between the other two patients | - |
| h | 2,7,128 | Maximum heart rate | > 146.9 BPM |

**Results**

In our project, the source data are obtained from a GE monitoring system spanning five intensive care units at Emory University Hospital. This system feeds a bedside electronic flowsheet (Cerner Millennium iView), where nurses enter data that go into the electronic health record, and a data archiving system, BedMasterEx (Excel Medical Corporation). BedMasterEx stores numeric data of heart rate, blood pressure, and oxygen saturation once a minute in a relational database. Waveform data are captured by BedMasterEx system at 240 Hz and stored in compressed binary files and links are recorded in the relational database.

Patient information including SBT start and end times is extracted from the Cerner iView system. This information is used to identify the waveform files to extract ECG and end-tidal $CO_2$ data for the required SBT interval plus an hour on either side. Instantaneous heart rate is calculated using ECG data after identifying RR intervals. Respiration rate is determined using end-tidal $CO_2$. Both instantaneous heart rate and respiration rate data help to correct approximately recorded start and end times of SBT using algorithms and also in clinical review. These data are deidentified for further analysis.

From these data, we the patients are grouped into 8 categories.

Table 3: Error of prediction for patient groups

| Group | Members | Error |
|---|---|---|
| Group a | 2,4,7,8,29,35,55,66,68,75,91,92,96,128 | 17.1% |

| Group b | 1,90,115,119,122 | 28.8% |
| Group c | 35,92 | 21.3% |
| Group d | 2,35,92,128 | 18.2% |
| Group e | 1,4,29,75,92,96,115,122 | 33.7% |
| Group f | 2,7,55,119 | 31.9% |
| Group g | 4,35,92 | 25.9% |
| Group h | 2,7,128 | 43.0% |

The response to the SBT for each patient category is show in figure 2.

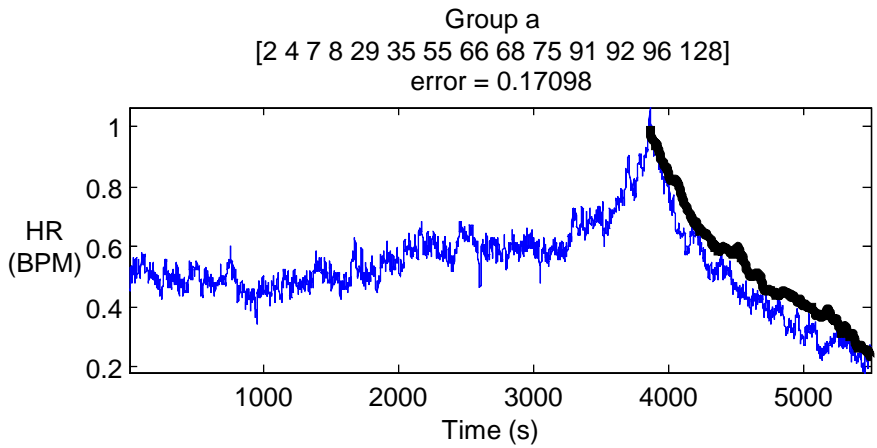

Group a
[2 4 7 8 29 35 55 66 68 75 91 92 96 128]
error = 0.17098

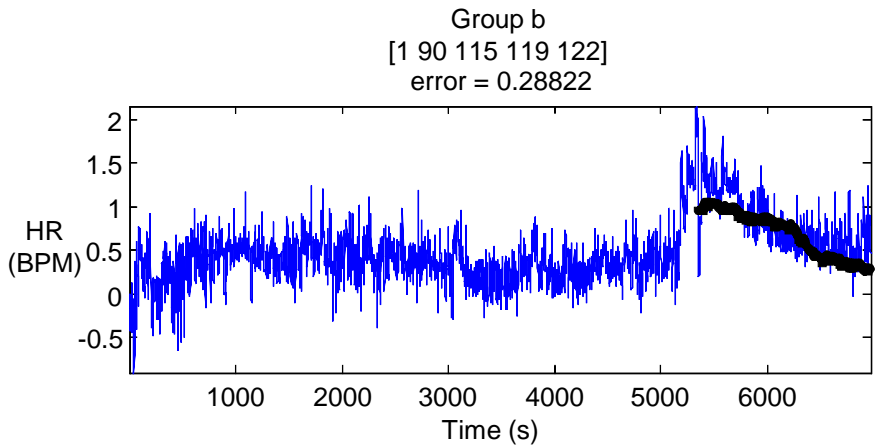

Group b
[1 90 115 119 122]
error = 0.28822

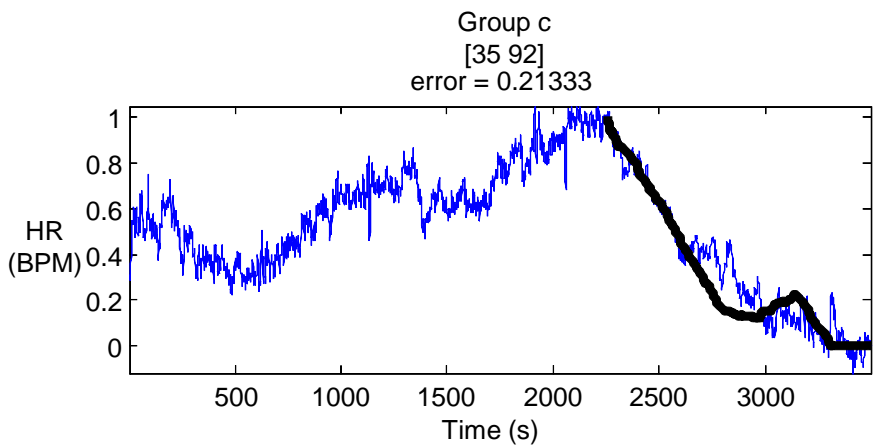

Group c
[35 92]
error = 0.21333

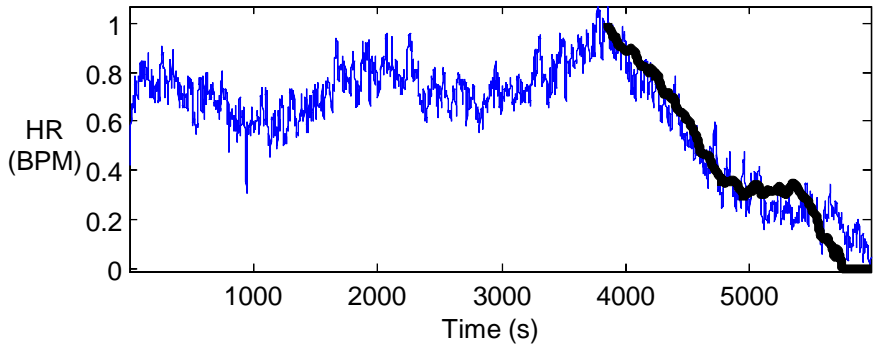

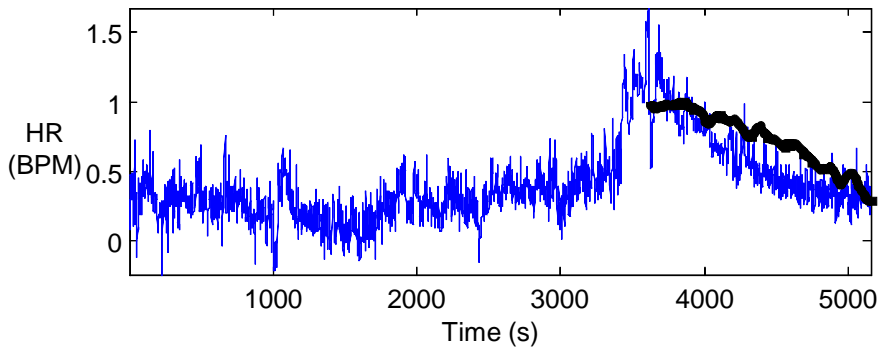

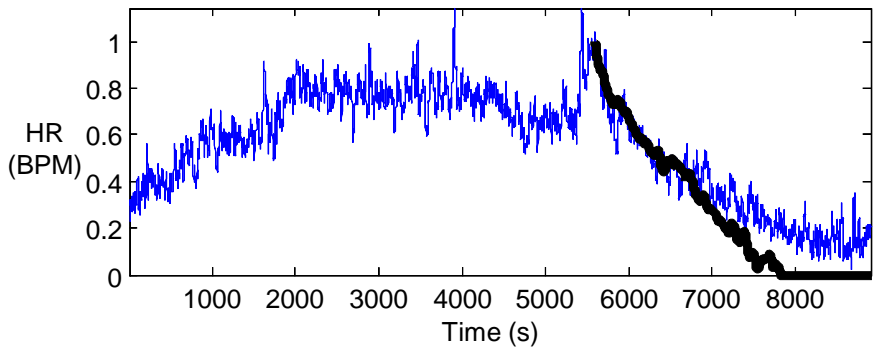

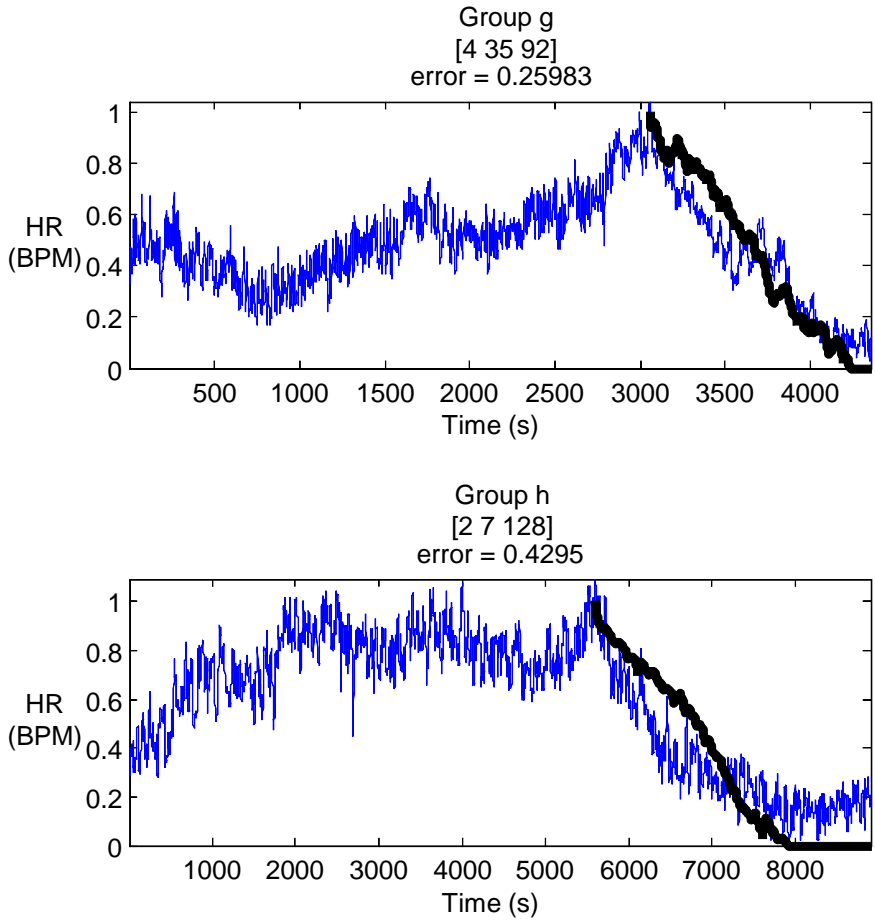

**Figure 2.** The HRR predictions (black) and average heart rates (blue) of patient groups (a)-(h) were obtained using the Full Cross Correlation; **(a)**, large mean HRR slopes; **(b)**, small mean HRR slope; **(c)**, patient pair with most similar mean HRR slope, maximum heart rate, and minimum heart rate; **(d)**, patient group with similar mean HRR slope; **(e)**, random assignments; **(f)**, random assignments; **(g)**, group c plus one additional patient whose mean HRR slope falls between the other two patients. **(h)**, maximum heart rate

**Discussion**
Visually, the predicted response to termination of the SBT is similar to the data for each of the seven patient groups. The prediction has an average error of 25%. The error is calculated using the group HRR prediction, $<\bar{f}>^{IIIp}$, and the group phase III average heart rate, $<\bar{f}_i>^{III}$:

$$\varepsilon = \frac{\left\| <\bar{f}>^{IIIp} - <\bar{f}_i>^{III} \right\|}{\left\| <\bar{f}_i>^{III} \right\|} \quad (11)$$

Here the $L_2$ norm is used.

Previously, we used a discrete-time, Gaussian, Markov model to predict the heart rate response to termination of the SBT [12]. While this approach produced many successful predictions, it had a few limitations. The error of predication in this patient-specific prediction ranges from 29% to 92% with average error rate around 66.9%. That approach could only produce an exponentially decaying response to termination of SBT. The more fully developed approach presented here has no restriction on the response function which it can produce, Eq. 4. For example, the previous approach could only predict concave responses, whereas the current approach can predict both concave and convex responses, see Figure 2.

Here, we predict the response for a group of patients. Previously, we had considered predicting the response to termination of SBT for each individual patient. Grouping of patients gives higher signal to noise ratio and is common clinical practice. The groupings show in Figure 2 lead to satisfactory predictive ability. Grouping of patients leads to a more accurate prediction of the heart rate in phase III. The reason for a lower error in the present prediction is that we utilized all the available fluctuations in Phase II in all the patients within a group with similar HRR profiles.

There are a number of practical issues when working with these SBT clinical data. The start time of phase II and III are recorded manually. This leads to an error of +/- 1 minute or so on these times. It is for this reason, that we use the average of heart rates around the start time of phase III as the initial value of the response to termination of SBT. The heart rates are noisy. Partly, this is due to intrinsic fluctuations in the heart rate of the patients, which are of interest to us and central to our analysis. In addition, however, there are fluctuations of the signal which are due, for example, to patient motion which disrupts the fidelity of the sensors. Less than 0.1% of the data are affected by this difficulty.

Equation 9 is general, for any grouping of the patients. Here, we show predictions for a few data sets. A large hospital system such as Emory is estimated to be able to provide ~15 data sets per day when all ICU beds are fully wired for data collection. With such a larger amount of data, design of the criteria for grouping patients becomes a critical issue. We expect factors such as patient age, sex, sedation level, average heart rate, and variance in heart rate will be effective criteria for grouping.

**Conclusion**

This novel marriage between the mathematics of fluctuations and clinical medicine contains a deeper postulate: the existence of what could be called "Newton's Laws for Biology." Modern biology embraces principles of scaling, modularity, and heritability. These principles speak to structure, distance, and space. Our proposed project, however, speaks to biological time, which is to the correlations and structure that exist in healthy, time-dependent physiology. We believe that there is an underlying structure to biological time and observables (the time series), which endows the healthy biological system with an ability to withstand stress and recover from trauma or illness. We have access to dense data, computational power, and far-from-equilibrium theory to explore biological time in specific and clinically important contexts.

While the proposed FDT approach clearly has important utility for cardiac physiology and for transitions between mechanical and free ventilation, it is anticipated that it will also be broadly applicable for predicting the response of patients to physiologic stress or clinical interventions, as it has in other complex systems [6, 10]. This applicability will be increasingly valuable as additional time-dependent critical care parameters become routinely monitored. Clinically, the importance lies in two realms. First, the clinical value of predicting what will happen "five minutes from right now" is key to safe management in acute care settings such as the emergency department, the operating room, and the ICU. Second, patients who veer away from their predicted physiology are "off trajectory" and constitute an immediate indication for heightened surveillance and early intervention. Early detection of such "off-trajectory" patients will likely accelerate detection of subtle pathological processes, which, left unchecked, would otherwise result in preventable complications, prolonged hospitalization and even death.